\documentclass[aps,pre,twocolumn,amsmath,amssymb]{revtex4}

\usepackage{graphicx}
\usepackage{dcolumn}
\usepackage{bm}

\begin{document}


\title{Time reversibility and nonequilibrium thermodynamics of second-order stochastic processes}

\author{Hao Ge}
\email{haoge@pku.edu.cn}
\affiliation{Beijing International Center for Mathematical Research (BICMR) and Biodynamic Optical Imaging Center (BIOPIC), Peking University, Beijing, 100871, PRC.}

\date{\today}

\begin{abstract}
Nonequilibrium thermodynamics of a general second-order stochastic system is investigated. We prove that at steady state, under inversion of velocities, the condition of time-reversibility over the phase space is equivalent to the antisymmetry of spatial flux and the symmetry of velocity flux. Then we show that the condition of time-reversibility alone could not always guarantee the Maxwell-Boltzmann distribution. Comparing the two conditions together, we found that the frictional force naturally emerges as the unique odd term of the total force at thermodynamic equilibrium, and is followed by the Einstein relation. The two conditions respectively correspond to two previously reported different entropy production rates. In case that the external force is only position-dependent, the two entropy production rates become one. We prove that such an entropy production rate can be decomposed into two nonnegative terms, expressed respectively by the conditional mean and variance of the thermodynamic force associated with the irreversible velocity flux at any given spatial coordinate. In the small inertia limit, the former term becomes the entropy production rate of the overdamped dynamics; while the anomalous entropy production rate originated from the latter term. Furthermore, regarding the connection between the First Law and Second Law, we found that in the steady state of such a limit, the anomalous entropy production rate is also the leading order of the Boltzmann-factor weighted difference between the spatial heat dissipation densities of the underdamped and overdamped dynamics, while their unweighted difference always tends to vanish.
\end{abstract}

\maketitle

\section{Introduction}

The stochastic movement of a macromolecule is usually described by Langevin equation, in which the Newtonian equation of motion is augmented by two additional force terms as consequences of thermal collisions with solvent molecules \cite{Langevin}. One term represents a frictional force and the other is a random fluctuating force, connecting to each other through Einstein relation \cite{Einstein}. When the frictional force dominates over inertial effects, the Langevin equation on the whole phase space could be further simplified into a first-order stochastic differential equation, i.e. overdamped dynamics, the corresponding time evolution of whose probability density function is given by the Smoluchowski equation \cite{Smoluchowski}. The very wide range of applications of the overdamped dynamics is also due to the fact that the microscopic and nanometric world of soft condensed matters is indeed at very low Reynolds numbers, which is the ratio of the inertial force and the viscous force \cite{Purcell}, guaranteeing the validity of the overdamped approximation for modeling cellular dynamics.

The overdamped dynamics without inertia is also an ideal model for studying nonequilibrium thermodynamics, which has been extensively studied for many years \cite{Onsager_Machlup,Sasa,JQQ,Seifert05,ep_fd_01,attila,phys_rep,esposito}. On the other hand, the condition of time-reversibility for second-order stochastic processes, which original Langevin equation belongs to, has already been derived \cite{Risken}, and two different corresponding entropy production rates were already put forward \cite{Qian_Kim,Ford}. Also, it was recently found that the overdamped approximation of Langevin equation fails to preserve the entropy production rate \cite{hidden_epr}, and an anomalous entropy production arises in the limit of small inertia.

Most of the previous works on the nonequilibrium thermodynamics of second-order stochastic processes were carried out from the perspective of fluctuation theorem, i.e. at the level of single trajectories \cite{Ford,hidden_epr}. However, what we can really measure is the fluxes of all physical quantities. Therefore, we turn to investigate from the perspective of fluxes \cite{Sc,JQQ,Hill}. In Sec. II, basic notations and the condition of time-reversibility for second-order stochastic processes are given (Eq. (\ref{condition2})), and the detailed balancing of spatial and velocity fluxes at steady state is emphasized (Eq. (\ref{detailed_balance})). In Sec. III, two simple examples with velocity-dependent external force are investigated, which are time-reversible but donot
satisfy the Maxwell-Boltzmann distribution. It further gives rise to a second condition for thermodynamic equilibrium beyond time-reversibility (Eq. (\ref{condition3})). In Sec. IV, we find out the two conditions correspond to two different entropy production rates respectively, and the entropy balance equation for one of them is related to the spatial densities of heat dissipation and mechanical work defined from the stochastic thermodynamics. The two entropy production rates become one in the absence of velocity-dependent external force, which will be studied in Sec. V. Then, a novel decomposition of entropy production rate is formulated in terms of probability fluxes in the phase space (Eq. (\ref{decom_epr})), and the leading orders of thermodynamic quantities in the limit of small inertia are calculated. We find that one term in the decomposition converges to the entropy production rate of the corresponding overdamped dynamics (Eq. (\ref{eqrspatial})); while the anomalous entropy production rate originates from the other term (Eq. (\ref{eqrano})). Furthermore, such a recently reported anomalous entropy production rate is also closely related to the steady-state flux of kinetic energy along the spatial coordinate (Eq. (\ref{333})).

\section{Time-reversibility and detailed balancing}

The meaning of time reversibility in physics, also known as the principle of microscopic reversibility \cite{Lewis,Onsager,Casimir}, is actually two folds: the forward trajectory $X(t)$ and its time-reversed one $X(-t)$ are both the solutions of a same deterministic microscopic equations of motion, or the statistical descriptions of the forward and backward processes are exactly the same \cite{Risken,Lewis}. In statistical mechanics, it is formulated in the second sense, and is equivalent to the principle of detailed balance \cite{Lewis,Tolman}.

In the present letter, we aim to study the general second-order stochastic differential equation in $n-$dimension describing the motion of a Brownian particle
\begin{equation}
m\frac{d^2\bm{X}}{dt^2}=\bm{F}(\bm{X},\bm{\dot{X}})+\xi(t),\label{Langevin_eq}
\end{equation}
which could be further rewritten as
\begin{eqnarray}
\frac{d\bm{X}}{dt}&=&\bm{V};\nonumber\\
m\frac{d\bm{V}}{dt}&=&\bm{F}(\bm{X},\bm{V})+\xi(t)
\end{eqnarray}
on the phase space, where $\xi(t)$ represents Gaussian white noise with position-dependent intensity
$D(\bm{x})$ and $m$ is the mass of the particle.

The time-reversibility of such a stochastic system should be formulated through the transition probabilities \cite{Seifert05,Tolman,Ford,Risken}, i.e.
\begin{eqnarray}
&&\mathbb{P}(\bm{X}(0)=\bm{x}_1,\bm{V}(0)=\bm{v}_1,\bm{X}(t)=\bm{x}_2,\bm{V}(t)=\bm{v}_2)\nonumber\\
&=&\mathbb{P}(\bm{X}(0)=\bm{x}_2,\bm{V}(0)=-\bm{v}_2,\bm{X}(t)=\bm{x}_1,\bm{V}(t)=-\bm{v}_1),\nonumber
\end{eqnarray}
for any positions $\bm{x}_1,\bm{x}_2$, velocities $\bm{v}_1,\bm{v}_2$ and time $t$.

It follows that
\begin{eqnarray}
&&\rho_0(\bm{x}_1,\bm{v}_1)p_t(\bm{x}_2,\bm{v}_2|\bm{x}_1,\bm{v}_1)\nonumber\\
&=&\rho_0(\bm{x}_2,-\bm{v}_2)p_t(\bm{x}_1,-\bm{v}_1|\bm{x}_2,-\bm{v}_2),\label{time_reversibility}
\end{eqnarray}
where $\rho_0(\bm{x},\bm{v})=\mathbb{P}(\bm{X}(0)=\bm{x},\bm{V}(0)=\bm{v})$ is the initial distribution and
\begin{eqnarray}
&&p_t(\bm{x},\bm{v}|\bm{\tilde{x}},\bm{\tilde{v}})\nonumber\\
&=&\mathbb{P}(\bm{X}(t)=\bm{x},\bm{V}(t)=\bm{v}|\bm{X}(0)=\bm{\tilde{x}},\bm{V}(0)=\bm{\tilde{v}})\nonumber
\end{eqnarray}
is the transition probability density function.

The time evolution of the function $p_t(\bm{x},\bm{v}|\bm{\tilde{x}},\bm{\tilde{v}})$ obeys both the Kolmogorov forward and backward equations:
\begin{eqnarray}
\frac{dp_t(\bm{\tilde{x}},\bm{\tilde{v}}|\bm{x},\bm{v})}{dt}&=&-\nabla_{\bm{\tilde{x}}}\cdot \left( \bm{\tilde{v}}p_t\right)-\nabla_{\bm{\tilde{v}}}\cdot \left( \frac{\bm{F}(\bm{\tilde{x}},\bm{\tilde{v}})}{m}p_t\right)\nonumber\\
&&+\frac{D(\bm{\tilde{x}})}{2m^2}\nabla_{\bm{\tilde{v}}}\cdot \nabla_{\bm{\tilde{v}}} p_t,\label{Kolmo_forward}\\
\frac{dp_t(\bm{\tilde{x}},\bm{\tilde{v}}|\bm{x},\bm{v})}{dt}&=&\bm{v}\cdot\nabla_{\bm{x}}p_t+\frac{\bm{F}(\bm{x},\bm{v})}{m}\cdot \nabla_{\bm{v}}p_t\nonumber\\
&&+\frac{D(\bm{x})}{2m^2}\nabla_{\bm{v}}\cdot \nabla_{\bm{v}} p_t.\label{Kolmo_backward}
\end{eqnarray}

Then we can apply the backward equation (\ref{Kolmo_backward}) for $p_t(\bm{x}_1,\bm{v}_1|\bm{x}_2,-\bm{v}_2)$ to derive the time evolution of function $$f_t(\bm{x}_2,\bm{v}_2,\bm{x}_1,\bm{v}_1)=\frac{\rho_0(\bm{x}_2,-\bm{v}_2)}{\rho_0(\bm{x}_1,\bm{v}_1)}p_t(\bm{x}_1,-\bm{v}_1|\bm{x}_2,-\bm{v}_2),$$
i.e.
\begin{eqnarray}
\frac{df_t}{dt}&=&-\nabla_{\bm{x}_2}\cdot \left( \bm{v}_2f_t\right)-\nabla_{\bm{v}_2}\cdot (\frac{\bm{F}^{rev}(\bm{x}_2,\bm{v}_2)}{m}f_t)\nonumber\\
&&+\frac{D(\bm{x}_2)}{2m^2}\nabla_{\bm{v}_2}\cdot \nabla_{\bm{v}_2} f_t+\bm{S}(\bm{x}_2,\bm{v}_2)\cdot f_t,\label{111}
\end{eqnarray}
where
$$\bm{F}^{rev}(\bm{x}_2,\bm{v}_2)=\bm{F}(\bm{x}_2,-\bm{v}_2)+\frac{D(\bm{x}_2)}{m}\nabla_{\bm{v}_2}\log \rho_0(\bm{x}_2,-\bm{v}_2)$$
and
\begin{eqnarray}
&&\bm{S}(\bm{x}_2,\bm{v}_2)\nonumber\\
&=&\left[ \nabla_{\bm{x}_2}\cdot \left(\bm{v}_2\rho_0(\bm{x}_2,-\bm{v}_2)\right)+\nabla_{\bm{v}_2}\cdot \left( \frac{\bm{F}(\bm{x}_2,-\bm{v}_2)}{m}\rho_0(\bm{x}_2,-\bm{v}_2)\right)\right.\nonumber\\
&&\left. +\frac{D(\bm{x}_2)}{2m^2}\nabla_{\bm{v}_2}\cdot \nabla_{\bm{v}_2} \rho_0(\bm{x}_2,-\bm{v}_2)\right]/\rho_0(\bm{x}_2,-\bm{v}_2).\nonumber
\end{eqnarray}

According to the definition of time-reversibility (Eq. (\ref{time_reversibility})), $f_t(\bm{x}_2,\bm{v}_2,\bm{x}_1,\bm{v}_1)\equiv p_t(\bm{x}_2,\bm{v}_2|\bm{x}_1,\bm{v}_1)$, which satisfies (\ref{Kolmo_forward}). Comparing (\ref{Kolmo_forward}) and (\ref{111}), we can get the necessary and sufficient condition for time-reversibility
\begin{eqnarray}
\bm{F}^{rev}(\bm{x}_2,\bm{v}_2)&=&\bm{F}(\bm{x}_2,\bm{v}_2);\nonumber\\
\bm{S}(\bm{x}_2,\bm{v}_2)&=&0.\label{condition1}
\end{eqnarray}

Define another function $\bm{\epsilon F}(\bm{x},\bm{v})$, which reverses all the odd variables including $\bm{v}$ in the function $\bm{F}(\bm{x},\bm{v})$ under the time reversal transformation, then the necessary and sufficient condition (\ref{condition1}) for time reversibility can be rewritten as \cite{Risken}
\begin{eqnarray}
\frac{D(\bm{x})}{m}\nabla_{\bm{v}}\log\rho_0&=&\bm{F}-\bm{\epsilon F};\nonumber\\
2D(\bm{x})\bm{v}\cdot\nabla_{\bm{x}}\log\rho_0&=&-\left[\bm{F}^2-(\bm{\epsilon F})^2\right.\nonumber\\
&&\left.+\frac{D(\bm{x})}{m}\nabla_{\bm{v}}\cdot(\bm{F}+\bm{\epsilon F})\right].\label{condition2}
\end{eqnarray}

The lower subequation in (\ref{condition2}) guarantees that the distribution $\rho_0$ is just the unique stationary solution $\rho^{ss}$ to the well-known Fokker-Planck equation of the probability distribution $\rho_t(\bm{x},\bm{v})$ in the phase space
\begin{equation}
\frac{\partial}{\partial t}\rho_t=-\nabla_{\bm{x}}\cdot \bm{J}_{\bm{x}}-\nabla_{\bm{v}}\cdot \bm{J}_{\bm{v}},\label{FP1}
\end{equation}
where the coordinate flux $\bm{J}_{\bm{x}}=\bm{v}\rho_t$ and the velocity flux $\bm{J}_{\bm{v}}=\frac{\bm{F}}{m}\rho_t-\frac{D(\bm{x})}{2m^2}\nabla_{\bm{v}}\rho_t$. Hence at steady state, time-reversibility is only equivalent to the upper subequation in (\ref{condition2}), just replacing $\rho_0$ with the unique stationary solution $\rho^{ss}$.

Furthermore, the time reversibility of second-order stochastic system at steady state also corresponds to certain symmetry of the probability fluxes in the phase space: the spatial fluxes are odd while the velocity fluxes are even with respect to all the odd variables including $\bm{v}$, i.e.
\begin{eqnarray}
\bm{j}_{\bm{x}}&=&-\bm{\epsilon j}_{\bm{x}};\nonumber\\
\bm{j}_{\bm{v}}&=&\bm{\epsilon j}_{\bm{v}}.\label{detailed_balance}
\end{eqnarray}
It could be regarded as {\bf the condition of detailed balance}. In steady state, it is equivalent to the condition of time-reversibility (just the upper subequation in (\ref{condition2})). It is quite different from that of the overdamped Langevin dynamics \cite{JQQ}, where all the fluxes vanish at time-reversibility.

\section{A second condition for thermodynamic equilibrium}

It is generally believed that time-reversibility of the statistical description for the stochastic process, or the detailed balance, is equivalent to the thermodynamic equilibrium, which has already been proved for chemical kinetics as well as overdamped dynamics \cite{Lewis,esposito,Seifert05,JQQ,ep_fd_01}. However, that might not be the case for the second-order stochastic dynamics. In case that there is certain velocity-dependent external force as a feedback control in addition to the frictional force, the application of overdamped approximation is totally invalid and the relation between time-reversibility and thermodynamic equilibrium has not been clear yet.

Indeed, a lot of forces $\bm{F}$ could satisfy the time-reversibility condition (\ref{condition2}) for certain given steady-state probability density $\rho^{ss}$. It is because once we have a well defined density function $\rho^{ss}$, then we could find an odd and an even function $\bm{f}_1(=-\bm{\epsilon f}_1)$ and $\bm{f}_2(=\bm{\epsilon f}_2)$, which are the solutions of the following equations that are derived from (\ref{condition2})
\begin{eqnarray}
\frac{D(\bm{x})}{m}\nabla_{\bm{v}}\log\rho^{ss}&=&\bm{f}_1;\nonumber\\
2m\bm{v}\cdot\nabla_{\bm{x}}\rho^{ss}&=&-\nabla_{\bm{v}}\cdot(\bm{f}_2\rho^{ss}).
\end{eqnarray}
The solution $\bm{f}_2$ is not even generally unique, since we can add any magnetic term $B\times\bm{v}$ to $\bm{f}_2$ in which $B$ is also odd under time reversal. Then $\bm{F}=\frac{\bm{f}_1+\bm{f}_2}{2}$ would make the second-order stochastic process (\ref{Langevin_eq}) become time-reversible.

The simplest example in which time-reversibility is not equivalent to thermodynamic equilibrium is when the velocity-dependent external force is linear, i.e.
 \begin{equation}
m\frac{d^2\bm{X}}{dt^2}=-\eta \bm{v}-\alpha \bm{v}-\nabla_{\bm{x}}U(\bm{x})+\xi(t),
\end{equation}
with certain positive constant $\alpha$. In this specific example, we assume that the diffusion coefficient as well as temperature is not position-dependent. Due to the Einstein relation $\eta=\frac{D}{2k_BT}$, where $T$ is the uniform temperature of the heat reservoir. Hence
the corresponding stationary distribution becomes $C\exp[-(\frac{1}{2}m\bm{v}^2+U(\bm{x}))/k_BT_{eff}]$ with the effective temperature $T_{eff}=\frac{\eta}{\eta+\alpha}T$.

The condition (\ref{condition2}) of time-reversibility is always satisfied in this example for all $\alpha \geq 0$. However, the effective temperature $T_{eff}=\frac{\eta}{\eta+\alpha}T$ is less than the temperature $T$ of the heat reservoir unless $\alpha=0$, which implies that it is not at thermal equilibrium with the reservoir. Such a nonequilibrium phenomenon has already been successfully used to
reduce the thermal fluctuations of a cantilever in atomic force microscopy(AFM) through feedback control \cite{Liang2000}, in which the force between the sharp tip at the end of the cantilever and the target surface obeys the Hooke's law and there is an external agent detecting the
velocity of the cantilever and performing the feedback.

Another nontrivial example is when $\rho^{ss}\propto \exp\left(-\frac{\frac{1}{2}m\bm{v}^2g(\bm{x})+U(\bm{x})}{k_BT}\right)$. In this case, we can set
$\bm{f}_1=-2\eta\bm{v}g(\bm{x})$ and any function $\bm{f}_2$ satisfying
$$\nabla_{\bm{v}}\cdot(\bm{f}_2\rho^{ss})=2m\bm{v}\cdot\left( \frac{m\bm{v}^2}{2k_BT}\nabla_{\bm{x}}g(\bm{x})+\frac{1}{k_BT}\nabla_{\bm{x}}U(\bm{x})\right)\rho^{ss},$$
then the second-order stochastic process (\ref{Langevin_eq}) with $\bm{F}=\frac{\bm{f}_1+\bm{f}_2}{2}$ would be time-reversible. One solution of $\bm{f}_2$ is
$$\bm{f}_2=-m\bm{v}^2\frac{\nabla_{\bm{x}} g(\bm{x})}{g(\bm{x})}-2k_BT\frac{\nabla_{\bm{x}} g(\bm{x})}{g(\bm{x})^2}-\frac{2\nabla_{\bm{x}}U(\bm{x})}{g(\bm{x})}.$$

Thus the condition of time-reversibility is only a necessary condition for thermodynamic equilibrium, because it alone could not be sufficient to guarantee the well-known Maxwell-Boltzmann distribution \cite{Maxwell,Boltzmann} in terms of the temperature of the reservoir.

We regard the equilibrium Maxwell-Boltzmann distribution as the second condition for thermodynamic equilibrium beyond time-reversibility for underdamped stochastic processes. This distribution
\begin{equation}
\rho^{eq}(\bm{x},\bm{v})=C(x)e^{-\frac{\frac{1}{2}m \bm{v}^2}{k_BT}}
\end{equation}
could be further rewritten as
\begin{eqnarray}
\frac{k_BT}{m}\nabla_{\bm{v}}\log\rho^{eq}=-\bm{v},\label{condition3}
\end{eqnarray}
in which the temperature $T$ of the reservoir should be position-independent.

On the other hand, if the steady-state distribution of the Fokker-Planck equation (\ref{FP1}) is in the Maxwell-Boltzmann type, i.e. $\rho^{ss}(\bm{x},\bm{v})=Ce^{-\frac{\frac{1}{2}m \bm{v}^2+U(\bm{x})}{k_BT}}$, then the total force $\bm{F}=-\eta \bm{v}-\nabla_{\bm{x}}U(x)+\bm{A}(\bm{x},\bm{v})$ where $\bm{A}$ must satisfy $\nabla_{\bm{v}}\cdot \left(\bm{A}\rho^{ss}\right)=0$. In order to make such a stochastic process be time-reversible, according to the condition (\ref{condition2}), we must further require the $\bm{A}(\bm{x},\bm{v})$ to be invariant under the inversion of velocity, e.g. when $\bm{A}(\bm{x},\bm{v})$ is the Lorentz force.

Therefore, a steady state is at thermodynamic equilibrium if and only if both (\ref{condition2}) and (\ref{condition3}) are satisfied, i.e.
\begin{eqnarray}
\frac{D(\bm{x})}{m}\nabla_{\bm{v}}\log\rho^{eq}=\bm{F}-\bm{\epsilon F}=-\frac{D(\bm{x})}{k_BT}\bm{v}.
\end{eqnarray}

The second equality implies that, at thermodynamic equilibrium, $\bm{F}$ could be naturally decomposed into an odd term $-\frac{D(\bm{x})}{2k_BT} \bm{v}$ and an even term $\frac{\bm{F}+\bm{\epsilon F}}{2}$. The unique odd term is a frictional force $-\eta(\bm{x})\bm{v}$ with $\eta(\bm{x}) =\frac{D(\bm{x})}{2k_BT}$, which is indeed the Einstein relation.

Hence in the following, we rewrite the force $\bm{F}$ as $-\eta(\bm{x})\bm{v}+\bm{F}_e$, in which $\bm{F}_e$ is regarded as the external force and we assume the local Einstein relation holds, i.e. $D(\bm{x})=2\eta(\bm{x})k_BT(\bm{x})$. Here we restrict ourselves to the case that the temperature at each instantaneous position of the corresponding particle is imposed just by the heat reservoir \cite{VanKampen,hidden_epr}, i.e. just replacing $T$ with $T(\bm{x})$ in the Langevin dynamics regardless with other complications \cite{Polettini}. Also we donot explicitly exclude the potential force from the external force $\bm{F}_e$, since such a convention would not affect any of the results below.

\section{Entropy production rates and the conservation of kinetic energy}

The problem of defining entropy production rate for underdamped stochastic processes has still not completely determined yet. Recently, a general definition of entropy production rate through comparison of the probabilities of the original path $\omega_{[0,T]}=\{\omega_s:0\leq s\leq T\}$ with respect to its time-reversed counterpart $r\omega_{[0,T]}=\{(r\omega)_s=\epsilon\omega_{T-s}:0\leq s\leq T\}$ has been put forward and already applied to stochastic systems with odd and even variables \cite{Seifert05,Ford}. Such an entropy production rate is expressed as
\begin{eqnarray}
&&epr\nonumber\\
&=&\lim_{T\rightarrow 0^+}\frac{1}{T}\left\langle\log\frac{P(\omega_{[t,t+T]})}{P(r\omega_{[t,t+T])}}\right\rangle_P\nonumber\\
&=&\int\int \frac{2}{D(\bm{x})}\left[\left(\frac{\bm{F}-\bm{\epsilon F}}{2}\right)-\frac{D(\bm{x})}{2m}\nabla_{\bm{v}}\log\rho_t\right]^2\rho_t d\bm{x}d\bm{v}.\nonumber
\end{eqnarray}

The entropy production rate $epr$ is nonnegative. According to the upper subequation in (\ref{condition2}), $epr=0$ is equivalent to the time-reversibility of the second-order stochastic system (\ref{Langevin_eq}) provided that it is already in steady state.

On the other hand, people have also defined another entropy production rate \cite{Qian_Kim}
\begin{eqnarray}
\widetilde{epr}=\int\int \frac{2}{D(\bm{x})}\left[-\eta(\bm{x})\bm{v}-\frac{D(\bm{x})}{2m}\nabla_{\bm{v}}\log\rho_t\right]^2\rho_t d\bm{x}d\bm{v}.\nonumber
\end{eqnarray}
This second entropy production rate $\widetilde{epr}$ is nonnegative and $\widetilde{epr}=0$ is equivalent to the condition of Maxwell-Boltzmann distribution (\ref{condition3}), which is usually not a consequence of (\ref{condition2}) as stated in Sec. III. An equilibrium state has to satisfy both of them, i.e. $\widetilde{epr}=0$ together with $epr=0$.

Such an entropy production rate $\widetilde{epr}$ is originally defined from the entropy balance equation \cite{Qian_Kim}
\begin{equation}
\frac{dS}{dt}=\widetilde{epr}-\widetilde{d_eS}_1-\widetilde{d_eS}_2,\label{entropy_balance2}
\end{equation}
where the entropy of the stochastic system in the whole phase space
\begin{eqnarray}
S=-\int\int \rho_t(\bm{x},\bm{v})\log \rho_t(\bm{x},\bm{v}) d\bm{x}d\bm{v},\nonumber
\end{eqnarray}
and the two additional terms
\begin{equation}
\widetilde{d_eS}_1=-\int\int\nabla_{\bm{v}}\cdot\left(\frac{\bm{F}(\bm{x},\bm{v})+\eta(\bm{x})\bm{v}}{m}\right)\rho_t d\bm{x}d\bm{v},\nonumber
\end{equation}
\begin{eqnarray}
\widetilde{d_eS}_2=\int\int \frac{\left(\eta(\bm{x})\bm{v}+\frac{D(\bm{x})}{2m}\nabla_{\bm{v}}\log\rho_t\right)\bm{v}\rho_t}{k_BT(\bm{x})}d\bm{v}d\bm{x}.\nonumber
\end{eqnarray}

The first additional term $\widetilde{d_eS}_1$ characterizes whether there is any velocity-dependent external force as feedback control on the system. It vanishes for any probability distribution $\rho_t$ on the phase space if and only if the external force $\bm{F}_e=\bm{F}+\eta(\bm{x})\bm{v}$ is independent of the velocity $\bm{v}$.

The second additional term $\widetilde{d_eS}_2$ can be further rewritten in terms of the spatial heat dissipation density $Q(\bm{x},t)=\int \left(\eta(\bm{x})\bm{v}+\frac{D(\bm{x})}{2m}\nabla_{\bm{v}}\log\rho_t\right)\bm{v}\rho_td\bm{v}$, i.e.
\begin{eqnarray}
\widetilde{d_eS}_2=\int\frac{Q(\bm{x},t)}{k_BT(\bm{x})}d\bm{x}.\nonumber
\end{eqnarray}

It is because in stochastic thermodynamics the dissipated heat into the heat bath could be defined as \cite{Seifert05,Qian_Kim}
\begin{eqnarray}
Q(t)&=&-\langle \left(-\eta(\bm{x})\bm{v}+\xi(t)\right)\circ d\bm{x}\rangle\nonumber\\
&=&\int \int \left(\eta(\bm{x})\bm{v}+\frac{D(\bm{x})}{2m}\nabla_{\bm{v}}\log\rho_t\right)\bm{v}\rho_td\bm{x}d\bm{v}\nonumber\\
&=&\int Q(\bm{x},t)d\bm{x}.
\end{eqnarray}

And the work $W(t)$ done by the external force $\bm{F}_e$ is
\begin{eqnarray}
W(t)&=&\langle \left(\bm{F}+\xi(t)\right)\circ d\bm{x}\rangle+Q(t)=\langle \bm{v}\circ md\bm{v}\rangle+Q(t)\nonumber\\
&=&\int \int \bm{F}_e\bm{v}\rho_td\bm{x}d\bm{v}=\int W(\bm{x},t)d\bm{x}\nonumber,
\end{eqnarray}
in which $W(\bm{x},t)=\int \bm{F}_e\bm{v}\rho_td\bm{v}$ is the spatial density of work done upon the system at position $\bm{x}$.

The products in the definition of heat and work along trajectory are both in the Stratonovich type. It is because only the Stratonovich type has the typical chain rule, which is needed for the validity of the First Law.

Furthermore, the spatial density of kinetic energy at position $\bm{x}$ is $E_t^{kinetic}(\bm{x})=\int \frac{1}{2}m\bm{v}^2\rho_t(\bm{x},\bm{v})d\bm{v}$, whose evolution involves the local densities of heat and work as well as the flow of kinetic energy along the spatial coordinate. Thus the energy conservation at position $\bm{x}$ becomes
$$\frac{d}{dt}E_t^{kinetic}(\bm{x})+\nabla_{\bm{x}}\cdot\bm{J}_{\bm{x}}^{kinetic}=W(\bm{x},t)-Q(\bm{x},t),$$
in which $\bm{J}_{\bm{x}}^{kinetic}=\int \frac{1}{2}m\bm{v}^2\bm{j}_{\bm{x}}d\bm{v}$ is the spatial flux of kinetic energy. In steady state, $W(\bm{x})-Q(\bm{x})=\nabla_{\bm{x}}\cdot\bm{J}_{\bm{x}}^{kinetic}$, which immediately implies the total $W^{ss}=Q^{ss}$. When thermodynamic equilibrium is reached, the spatial densities of heat dissipation and work as well as the spatial flux of kinetic energy all vanish.

For instance, in the steady state of the process (\ref{Langevin_eq}) when $\bm{F}=-\eta \bm{v}-\alpha \bm{v}-\nabla_{\bm{x}}U(\bm{x})$, we have $\widetilde{epr}=n\frac{\alpha^2}{m(\eta+\alpha)}$, $\widetilde{d_eS}_1=n\frac{\alpha}{m}$ and $\widetilde{d_eS}_2=-\frac{n\alpha\eta}{m(\alpha+\eta)}$ \cite{Qian_Kim}. When $\alpha$ is positive, $\widetilde{d_eS}_1>0$ and the total heat dissipation $Q^{ss}=k_BT\widetilde{d_eS}_2<0$, which indicates that the heat flows from the reservoir to the internal system and then is released to the external agent in the form of mechanical work, sustaining such a nonequilibrium steady state \cite{Qian_Kim}. It is because the mechanical work done by the extrnal force $-\alpha\bm{v}$ is always negative, which must be balanced by the heat input into the internal system.

\section{In the absence of velocity-dependent external force}

\subsection{The condition of equilibrium state}

When $\bm{F}=-\eta(\bm{x}) \bm{v}+G(\bm{x})$ where the external force $G(\bm{x})$ is only dependent on position, the two entropy production rates $epr$ and $\widetilde{epr}$ become one.

In this case, the condition of time reversibility (\ref{condition2}) is already equivalent to thermodynamic equilibrium, and could be simply rewritten as
\begin{eqnarray}
k_BT(\bm{x})\nabla_{\bm{v}}\log\rho^{eq}&=&-m\bm{v};\nonumber\\
k_BT(\bm{x})\bm{v}\cdot\nabla_{\bm{x}}\log\rho^{eq}&=&-\bm{v}\cdot G(\bm{x}).\label{eq19}
\end{eqnarray}

The upper subequation guarantees the equilibrium distribution to be at the form of $C(\bm{x})e^{-\frac{m\bm{v}^2}{2k_BT(\bm{x})}}$. Then the lower subequation further requires
$$k_BT(\bm{x})\bm{v}\cdot\nabla_{\bm{x}}\left (\log C(\bm{x})-\frac{m\bm{v}^2}{2k_BT(\bm{x})}\right)=-\bm{v}\cdot G(\bm{x})$$
to be hold for any coordinates $\bm{x}$ and $\bm{v}$. Hence both the coefficients in front of $\bm{v}$ and $\bm{v}^3$ have to vanish. Therefore (\ref{eq19}) holds if and only if the external force $G(\bm{x})$ associates with a potential and the temperature profile $T(\bm{x})$ imposed from the heat reservoir is independent of $\bm{x}$. In such an equilibrium case, the Fokker-Planck equation of Langevin dynamics is already known as the Kramers' equation, which takes the Maxwell-Boltzmann distribution as the corresponding equilibrium distribution \cite{Kramers}.


\subsection{The decomposition of entropy production rate}

As we already know, the velocity flux $\bm{j}_{\bm{v}}$ can be decomposed into an irreversible term $\bm{j}^{ir}_{\bm{v}}=\left(-\frac{\eta(\bm{x})}{m}\bm{v}-\frac{D(\bm{x})}{2m^2}\nabla_{\bm{v}}\log\rho_t\right)\rho_t$ and a reversible term $\bm{j}^{rev}_{\bm{v}}=\frac{G(\bm{x})}{m}\rho_t$ \cite{Ford,Risken}. Then $\bm{f}^{ir}_{\bm{v}}=\frac{\bm{j}^{ir}_{\bm{v}}}{\rho_t}$ could be regarded as the {\em thermodynamic force} associated with the irreversible velocity flux $\bm{j}^{ir}_{\bm{v}}$. Hence the entropy production rate is just the average of the conditional second moment of $\bm{f}^{ir}_{\bm{v}}$ at given position, i.e.
\begin{eqnarray}
epr&=\int\frac{2m^2}{D(\bm{x})}\langle \left(\bm{f}^{ir}_{\bm{v}}\right)^2\rangle_{\bm{x}}\hat{\rho}_t(\bm{x})d\bm{x},\nonumber
\end{eqnarray}
where
$$\langle \left(\bm{f}^{ir}_{\bm{v}}\right)^2\rangle_{\bm{x}}=\int \left(\bm{f}^{ir}_{\bm{v}}\right)^2\frac{\rho_t}{\hat{\rho}_t(\bm{x})} d\bm{v}$$
is the second moment of $\bm{f}^{ir}_{\bm{v}}$ given $\bm{x}$,
and $\hat{\rho}_t(\bm{x})=\int \rho_t d\bm{v}$ is the marginal distribution of the spatial coordinate.

We notice the fact that
$$\langle \left(\bm{f}^{ir}_{\bm{v}}\right)^2\rangle_{\bm{x}}=\langle \bm{f}^{ir}_{\bm{v}}\rangle^2_{\bm{x}}+var_{\bm{x}}(\bm{f}^{ir}_{\bm{v}}),$$
where
$$\langle \bm{f}^{ir}_{\bm{v}}\rangle_{\bm{x}}=-\frac{\eta(\bm{x})}{m}\bm{J}_{\bm{x}}/\hat{\rho}_t(\bm{x})$$
is the mean and
$var_{\bm{x}}(\bm{f}^{ir}_{\bm{v}})$ is the variance of $\bm{f}^{ir}_{\bm{v}}$ at any given position $\bm{x}$. $\bm{J}_{\bm{x}}$ is the integrated spatial fluxes $\bm{J}_{\bm{x}}=\int \bm{j}_{\bm{x}}d\bm{v}$.

Then we can decompose the entropy production rate into two nonnegative terms
\begin{equation}
epr=epr^{spatial}+\Xi, \label{decom_epr}
\end{equation}
where
\begin{eqnarray}
epr^{spatial}&=&\int\frac{2m^2}{D(\bm{x})}\langle \bm{f}^{ir}_{\bm{v}}\rangle^2_{\bm{x}}\hat{\rho}_t(\bm{x})d\bm{x}\nonumber\\
&=&\int\frac{\eta(\bm{x})}{k_BT(\bm{x})}\frac{\bm{J}_{\bm{x}}^2}{\hat{\rho}_t(\bm{x})}d\bm{x}
\end{eqnarray}
is the spatial entropy production rate only regarding the integrated spatial flux, and
\begin{equation}
\Xi=\int\frac{2m^2}{D(\bm{x})}var_{\bm{x}}(\bm{f}^{ir}_{\bm{v}})\hat{\rho}_t(\bm{x})d\bm{x}.
\end{equation}


\subsection{In the limit of small inertia}

In order to investigate the thermodynamic behavior in the limit of small inertia, we need to decompose the right-hand-side operator of the Fokker-Planck equation (\ref{FP1}) into two terms, i.e.

$$\frac{\partial}{\partial t}\rho_t=\left[\eta(\bm{x})\mathcal{L}_1+\mathcal{L}_2\right]\rho_t,$$
in which
$$\mathcal{L}_1\rho_t=\frac{1}{m}\nabla_{\bm{v}}\cdot\left(\bm{v}\rho_t\right)+\frac{k_BT(\bm{x})}{m^2}\nabla_{\bm{v}}\cdot\nabla_{\bm{v}}\rho_t,$$
and
$$\mathcal{L}_2\rho_t=-\bm{v}\cdot\nabla_{\bm{x}}\rho_t-\frac{G(\bm{x})}{m}\nabla_{\bm{v}}\rho_t.$$

Denote $w(\bm{v}|\bm{x})$ as the locally approximated Maxwell-Boltzmann distribution $w(\bm{v}|\bm{x})=\frac{1}{\left(2\pi\frac{k_BT(\bm{x})}{m}\right)^{n/2}}e^{-\frac{m\bm{v}^2}{2k_BT(\bm{x})}}$, and a projection operator $\mathcal{P}$ is defined as
$$\left(\mathcal{P}f\right)(\bm{x},\bm{v})=w(\bm{v}|\bm{x})\int f(\bm{x},\bm{v})d\bm{v}.$$

Then borrowing the operator method in \cite{Gardiner85}, we can get the limiting equation satisfied by $\left(\mathcal{P}\rho_t\right)(\bm{x},\bm{v})=w(\bm{v}|\bm{x})\hat{\rho}_t(\bm{x})$ as
$$\frac{\partial}{\partial t}\left(\mathcal{P}\rho_t\right)=-\mathcal{P}\mathcal{L}_2\left(\eta(\bm{x})\mathcal{L}_1\right)^{-1}\mathcal{L}_2\left(\mathcal{P}\rho_t\right).$$
And the residual term
$$\left(1-\mathcal{P}\right)\rho_t\approx -\left(\eta(\bm{x})\mathcal{L}_1\right)^{-1}\mathcal{L}_2\left(\mathcal{P}\rho_t\right).$$

Using the eigenvalues and eigenfunctions of the operator $\mathcal{L}_1$, we can get that in the limit of small inertia, $\hat{\rho}_t(\bm{x})$ would satisfy the corresponding Smoluchowski equation \cite{Gardiner85,hidden_epr,anti-Ito}:
\begin{equation}
\frac{\partial \hat{\rho}_t(\bm{x})}{\partial t}=-\nabla_{\bm{x}}\cdot\bm{J}_{\bm{x}}^{over},\label{Fokker-Planck-Smol}
\end{equation}
in which the overdamped spatial flux $\bm{J}_{\bm{x}}^{over}=\frac{G(\bm{x})}{\eta(\bm{x})} \hat{\rho}_t(\bm{x})-\frac{1}{\eta (\bm{x})}\nabla_{\bm{x}}\left[k_BT(\bm{x})\hat{\rho}_t(\bm{x})\right]$.
Note that the frictional coefficient in the correct Smoluchowski equation, also known as Fokker-Planck equation, should be formulated in the anti-Ito form \cite{anti-Ito}.

We can further get the leading order for the residual term, and finally get
\begin{eqnarray}
&&\rho_t(\bm{x},\bm{v})\nonumber\\
&=&\hat{\rho}_t(\bm{x})w(\bm{v}|\bm{x})+w(\bm{v}|\bm{x})\frac{m\bm{v}}{k_BT(\bm{x})}\cdot\bm{J}_{\bm{x}}^{over}\nonumber\\
&&+\hat{\rho}_t(\bm{x})w(\bm{v}|\bm{x})\frac{m\bm{v}\cdot\nabla_{\bm{x}}T(\bm{x})}{\eta(\bm{x})k_BT^2(\bm{x})}\left[\frac{n+2}{6}k_BT(\bm{x})-\frac{m\bm{v}^2}{6}\right].\nonumber\\
\label{app_dist}
\end{eqnarray}

Eq. (\ref{app_dist}) implies the integrated spatial flux $\bm{J}_{\bm{x}}$ of the second-order stochastic system converges to the overdamped spatial flux $\bm{J}_{\bm{x}}^{over}$.
Therefore,
\begin{equation}
epr^{spatial}\approx epr^{over}=\int\frac{\eta(\bm{x})}{k_BT(\bm{x})}\left(\frac{\bm{J}^{over}_{\bm{x}}}{\hat{\rho}_t}\right)^2\hat{\rho}_td\bm{x},\label{eqrspatial}
\end{equation}
which is just the entropy production rate defined for the overdamped dynamics associated with (\ref{Fokker-Planck-Smol})\cite{JQQ}.

\begin{figure*}[ht]
\includegraphics[width=3in]{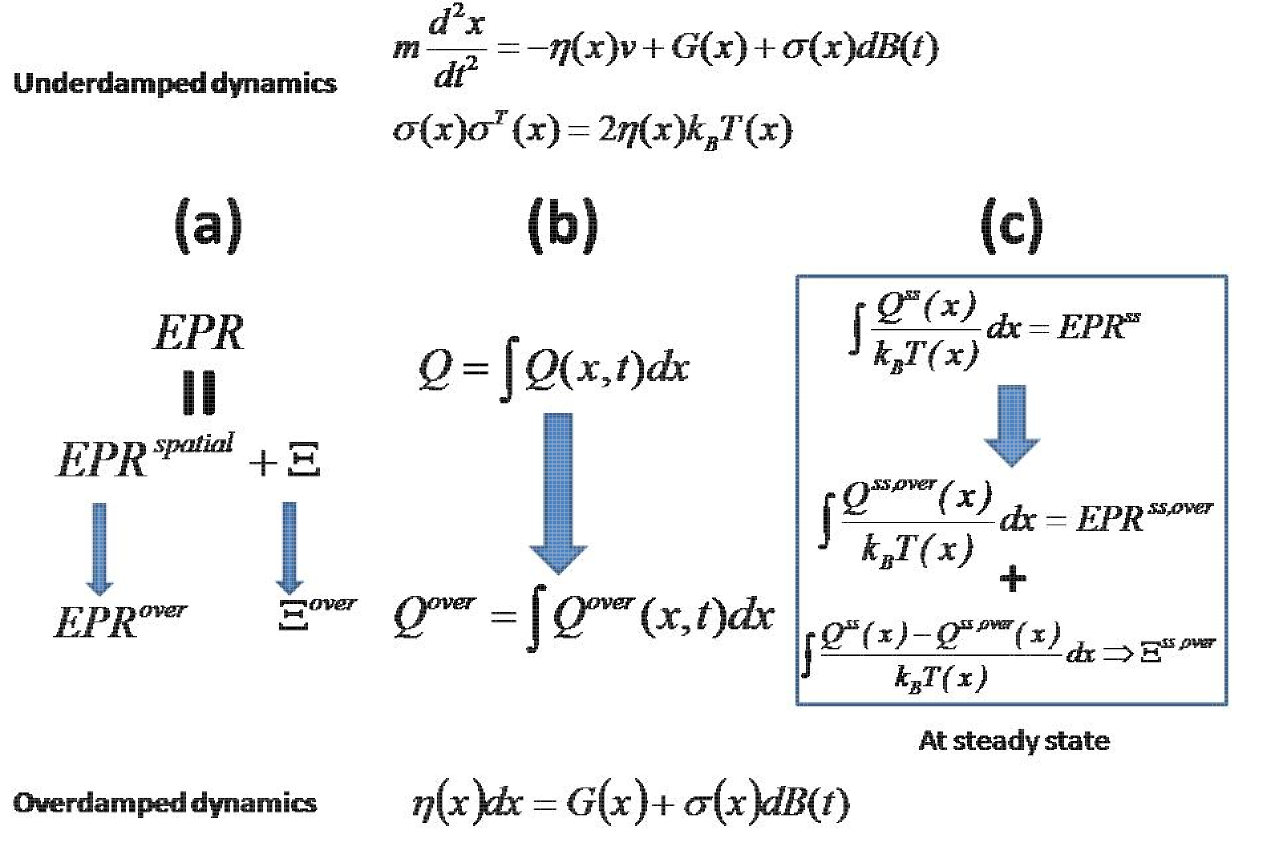}
\caption[fig_diagram]{Summary of the thermodynamic connection between underdamped stochastic dynamics and its corresponding overdamped dynamics. The diffusion term in the overdamped dynamics written here should be in the anti-Ito format, i.e. the Fokker-Planck equation of which is Eq. (\ref{Fokker-Planck-Smol}).} \label{fig_summary}
\end{figure*}

Further from (\ref{app_dist}), we can also have
$$var_{\bm{x}}(\bm{f}^{ir}_{\bm{v}})\approx\frac{n+2}{24}\left[\frac{D(\bm{x})}{\eta(\bm{x})m}\left(\frac{\nabla_{\bm{x}}T(\bm{x})}{T(\bm{x})}\right)\right]^2,$$
and finally
\begin{equation}
\Xi\approx \Xi^{over}=\frac{n+2}{6}k_B\int\frac{\left[\nabla_{\bm{x}}T(\bm{x})\right]^2}{\eta(\bm{x})T(\bm{x})}\hat{\rho}_t d\bm{x},\label{eqrano}
\end{equation}
in which $n$ is the dimension. It is the same as the anomalous entropy production rate from the perspective of single trajectories \cite{hidden_epr}. It implies the general decomposition of entropy production rate (\ref{decom_epr}) is consistent with that reported recently in the limit of small inertia (Fig. \ref{fig_summary}A) \cite{hidden_epr}. Thermodynamic equilibrium of the second-order stochastic system is equivalent to $epr^{over}=\Xi^{over}=0$.

Regarding the First Law of Thermodynamics, we found that the spatial densities of kinetic energy and work in the small inertia limit are approximated by
\begin{eqnarray}
&&E_t^{kinetic}(\bm{x})\approx \frac{n}{2}k_BT(\bm{x})\hat{\rho}_t,\nonumber\\
&&W(\bm{x})=G(\bm{x})\cdot\bm{J}_{\bm{x}}\approx G(\bm{x})\cdot\bm{J}_{\bm{x}}^{over},
\end{eqnarray}
and the flux of the kinetic energy
\begin{eqnarray}
\bm{J}_{\bm{x}}^{kinetic}&\approx&\frac{1}{2}(n+2)k_BT(\bm{x})\bm{J}_{\bm{x}}^{over}\nonumber\\
&&-\frac{n+2}{6}\frac{k_B^2T(\bm{x})}{\eta(\bm{x})}\left[\nabla_{\bm{x}}T(\bm{x})\right]\hat{\rho}_t(\bm{x}),\label{222}
\end{eqnarray}
in which the first term is the flux of kinetic energy associated with the flux of particle, and the second term is purely due to temperature gradient. In addition, we note that in the absence of external force, i.e. $G(\bm{x})=0$, then $\bm{J}_{\bm{x}}^{ss,over}=0$ and $\hat{\rho}^{ss}(\bm{x})\propto\frac{1}{T(\bm{x})}$. Hence due to (\ref{222}) the heat flux
$$\bm{J}_{\bm{x}}^{ss,kinetic}\propto -\frac{1}{\eta(\bm{x})}\nabla_{\bm{x}}T(\bm{x}),$$
which is consistent with the Fourier's law of heat conduction.

Then in the limit of small inertia, the spatial heat dissipation density
\begin{eqnarray}
Q(\bm{x},t)&\approx&W(\bm{x},t)-\nabla_{\bm{x}}\cdot\bm{J}_{\bm{x}}^{kinetic}-\frac{d}{dt}E_t^{kinetic}(\bm{x})\nonumber\\
&=&\left[G(\bm{x})-\frac{n}{2}k_B\nabla_{\bm{x}}T(\bm{x})\right]\cdot\bm{J}_{\bm{x}}^{over}\nonumber\\
&&-\nabla_{\bm{x}}\cdot\left[k_BT(\bm{x})\bm{J}_{\bm{x}}^{over}\right]\nonumber\\
&&+\frac{n+2}{6}\nabla_{\bm{x}}\cdot\left[\frac{k_BT(\bm{x})\nabla_{\bm{x}}k_BT(\bm{x})}{\eta(\bm{x})}\hat{\rho}_t(\bm{x})\right].\nonumber
\end{eqnarray}

On the other hand, the heat dissipation rate of the overdamped dynamics is defined as \cite{anti-Ito}
\begin{eqnarray}
Q^{over}(t)&=&\left\langle \left[G(\bm{x})-\frac{n}{2}k_B\nabla_{\bm{x}}T(\bm{x})\right]\circ d\bm{x}\right\rangle\nonumber\\
&=&\int \left[G(\bm{x})-\frac{n}{2}k_B\nabla_{\bm{x}}T(\bm{x})\right]\cdot\bm{J}_{\bm{x}}^{over}d\bm{x},\nonumber\\
&=&\int Q^{over}(\bm{x},t)d\bm{x},
\end{eqnarray}
where $Q^{over}(\bm{x},t)=\left[G(\bm{x})-\frac{n}{2}k_B\nabla_{\bm{x}}T(\bm{x})\right]\cdot\bm{J}_{\bm{x}}^{over}$ is the corresponding spatial density. Note that $Q^{over}(t)$ is just the leading order of $Q(t)$ in the limit of small inertia (Fig. \ref{fig_summary}B).

At the steady state of the second-order stochastic process, the spatial density of heat dissipation is directly related to the entropy production rate through the temperature profile (Fig. \ref{fig_summary}C), i.e.
$$\int\frac{Q^{ss}(\bm{x})}{k_BT(\bm{x})}d\bm{x}=epr^{ss}\geq 0,$$
where the Second Law stands.

Meanwhile, the Boltzmann factor weighted work converges to the entropy production rate of the overdamped dynamics
$$\int\frac{W^{ss}(\bm{x})}{k_BT(\bm{x})}d\bm{x}\approx \int\frac{Q^{ss,over}(\bm{x})}{k_BT(\bm{x})}d\bm{x}=epr^{ss,over}.$$
Hence the weighted difference between the heat dissipation densities of the original second-order process and its overdamped limit is
\begin{eqnarray}
\int\frac{Q^{ss}(\bm{x})-Q^{ss,over}(\bm{x})}{k_BT(\bm{x})}d\bm{x}&\approx &-\int\frac{\nabla_{\bm{x}}\cdot\bm{J}_{\bm{x}}^{ss,kinetic}}{k_BT(\bm{x})}d\bm{x}\nonumber\\
&\approx& \Xi^{ss,over},\label{333}
\end{eqnarray}
while the unweighted difference tends to vanish, i.e. $\int\left[Q^{ss}(\bm{x})-Q^{ss,over}(\bm{x})\right]d\bm{x}\approx 0$.

Until now, we have established the thermodynamic connection between underdamped stochastic dynamics and its corresponding overdamped dynamics (Fig. \ref{fig_summary}), concerning both the second law and first law.

\section{Conclusion and discussion}

 We have carefully studied nonequilibrium thermodynamics of the second-order stochastic system in terms of probability fluxes as well as the conservation of kinetic energy. For the Second Law, it stimulated a novel decomposition of entropy production rate, one term of which directly contacts the entropy production rate of the corresponding overdamped dynamics; for the First Law, the other term of the entropy production rate at steady state is associated with the Boltzmann factor weighted difference between the spatial heat dissipation densities of the underdamped and overdamped dynamics. In the present letter, we assumed that all the probabilities and fluxes vanish at infinity. The results here may need to be modified if the second-order stochastic system is confined in a finite domain with certain boundary conditions, and surely different boundary conditions should give rise to different modifications. These could be the subjects of further studies.

We thank Prof. Hong Qian and Prof. Min Qian for helpful discussions. We acknowledge the financial support from the Foundation for the Author
of National Excellent Doctoral Dissertation of China (No. 201119) and NSFC(No. 21373021).

\end{document}